\documentclass[12pt,preprint]{aastex}

\shorttitle{Understanding Simulations by Energy Equation}
\shortauthors{Lin et al.}

\begin{document}

\title{Understanding Simulations of Thin Accretion Disks by Energy Equation}

\author{Da-Bin Lin\altaffilmark{1},
Wei-Min Gu\altaffilmark{1,2},
Tong Liu\altaffilmark{1},
Mou-Yuan Sun\altaffilmark{1},
and Ju-Fu Lu\altaffilmark{1}}

\altaffiltext{1}{Department of Physics and Institute of Theoretical Physics
and Astrophysics, Xiamen University, Xiamen, Fujian 361005, China;
dabinlin@xmu.edu.cn, lujf@xmu.edu.cn}
\altaffiltext{2}{Harvard-Smithsonian Center for Astrophysics,
60 Garden Street, Cambridge, MA 02138, USA}

\begin{abstract}
We study the fluctuations of standard thin accretion disks by linear analysis
of the time-dependent energy equation together with the vertical hydrostatic
equilibrium and the equation of state.
We show that some of the simulation results in \citet{Hirose2009b},
such as the time delay, the relationship of power spectra,
and the correlation between magnetic energy and radiation energy,
can be well understood by our analytic results.
\end{abstract}


\section{Introduction}

Aperiodic X-ray fluctuations have been observed
from both galactic black hole binaries (BHBs) and active
galactic nuclei (AGNs) (Uttley et al. 2005; McHardy et al. 2006). 
The Power Spectral Density (PSD) of such variability is generally
modeled with a power law, $P(f) \propto f^{-\beta}$, where $P(f)$
is the power at frequency $f$, and ${\beta}$ varies with frequency. 
In the soft state, the PSDs of both BHBs and AGNs have 
a steep slope with $\beta \sim 2$ at high frequencies, 
flatting to a shallow slope with $\beta \sim 1$ below
a bend frequency $f_{\rm b}$, which is typically around $10\rm Hz$ for BHBs
(see \citealp{King2004} and references therein).
The PSDs in the hard state are more complex.
The origin of variability is not well understood yet.
However, it is highly tempting to relate this variability to the
magnetohydrodynamic (MHD) turbulence, which is believed to
drive the accretion process (\citealp{Balbus1991}).
Some works followed this path through numerical simulation,
and typically used proxies for the radiation 
rather than a direct measure of luminosity
(\citealp{HawleyK2001}; \citealp{Noble2009}).
It remains uncertain 
whether the proxies for radiation are appropriate 
to describe the luminosity fluctuations. 
In the present work, we will show that the variability 
of magnetic energy (stress) of the standard thin disk
is different from that of radiation 
for short time-scale fluctuations. 

Recently, shearing box simulations of stratified magnetorotational turbulence 
(\citealp{Hirose2009b}) showed that fluctuations in the magnetic energy (stress)
lead those in the radiation energy with roughly a thermal time-scale,
and a correlation is found between the stress and total pressure.
Moreover, the disk is found to be thermally stable, which is, however,
in conflict with the disk theory
(\citealp{Lightman1974}; \citealp{Shakura1976}).
The discrepancy reveals that the correlation found in the simulation 
may be different from the $\alpha$-prescription.
For example, such a correlation may be related to the energy equation
or result from the feedback from pressure to stress, which is not
in the form of the standard $\alpha$-prescription (e.g., \citealp{Lin2011}; \citealp{Ciesielski2012}).
Since the dissipation of magnetic energy will heat the gas of accretion flow,
the perturbations in the magnetic energy will produce corresponding 
fluctuations in the internal energy and therefore in the pressure.
Then, there should exist a correlation and delay between the stress
and the pressure. In the present work, based on the energy equation,
we will investigate the relationship between fluctuations of the viscous
heating and the inducing fluctuations of the radiative cooling.

The paper is organized as follows.
The relationship of fluctuations of the viscous heating and the inducing
fluctuations of the radiative cooling is derived in Section~2. 
A comparison of analysis and simulation is presented in Section~3.
Conclusions and discussion are made in Section~4.

\section{Radiative cooling fluctuations induced by viscous heating
fluctuations}

\subsection{Energy equation}

In the context of standard thin accretion disk (\citealp{Shakura1973}), 
the vertically integrated energy equation
in cylindrical coordinates ($r$, $\phi $, $z)$ takes the form
(e.g., \citealp{Kato2008}):
\begin{equation}\label{EEnergyInitial}
\frac{{\partial}E}{{\partial}t}
- \left(E+\Pi \right)\frac{{\partial}\ln\Sigma}{{\partial}t}
+ \Pi\frac{{\partial}{\ln}H}{{\partial}t}
= Q_{\rm {vis}}^+ - Q_{\rm {rad}}^- \ ,
\end{equation}
where $H$ is the vertical height of the disk, and ${\Sigma}\,(=2\rho H)$ and
$\Pi(=2pH)$ are the surface density and the vertically integrated pressure,
respectively. The gas internal energy $E$ and the radiative cooling rate
$Q_{\rm rad}^-$ per unit area are expressed as 
\begin{equation}\label{EEnergyE}
E=E_{\rm rad}+E_{\rm gas}= \left [ 3(1-\beta)+\frac{\beta}{\gamma-1} \right ] \Pi \ ,
\end{equation}
\begin{equation}\label{ERadiationCooling}
Q_{\rm {rad}}^- = \frac{16acT^4}{3\bar{\kappa}{\Sigma}} \ ,
\end{equation}
where $E_{\rm rad}$ and $E_{\rm gas}$ are respectively the internal energy
of radiation and gas,
$\beta$ is defined as the ratio of the gas to the total pressure,
i.e., $\beta \equiv \Pi_{\rm gas}/\Pi$,
$\gamma$ is the ratio of specific heating, and $T$ is the 
temperature on the equatorial plane of the disk. 
The opacity $\bar{\kappa}$ is generally dominated by the electron
scattering ($\kappa_{\rm es}$) in radiation-pressure-dominated accretion disks,
where $\bar{\kappa}$ can be regarded as a constant.
On the other hand, if the opacity is dominated by the free-free absorption ($\kappa_{\rm ff}$), 
$\bar{\kappa}$ will vary with the temperature and the density. 
The viscous heating $Q_{\rm vis}^+$ is due to the dissipation
of magnetic energy and turbulent kinetic energy in magnetoturbulent disks, 
and is dominated by the dissipation of magnetic energy in simulations
(\citealp{Simon2009}).

For the fluctuations with a time-scale less than the viscous time-scale,
the variation of $\Sigma$ can be neglected,
and Equation~(\ref{EEnergyInitial}) is therefore simplified as
\begin{equation}\label{EEnergy}
\frac{{\partial}E}{{\partial}t}
+ \Pi\frac{{\partial}{\ln}H}{{\partial}t}
= Q_{\rm {vis}}^+  - Q_{\rm {rad}}^-.
\end{equation}
In order to study the induced fluctuations of radiative cooling,
we adopt the vertical hydrostatic equilibrium
\begin{equation}\label{EVertical}
\Omega_{\rm K}^2H^2 = \frac{\Pi}{\Sigma},
\end{equation}
and the equation of state, which can be approximately expressed as
\begin{equation}\label{EState}
\Pi = \Pi_{\rm {gas}}+\Pi_{\rm {rad}} = \frac{k_{\rm B}}
{\mu m_{\rm H}}{\Sigma}T + \frac{2}{3}aT^4H.
\end{equation}

\subsection{Relationship of fluctuations}
\begin{figure}
\plotone{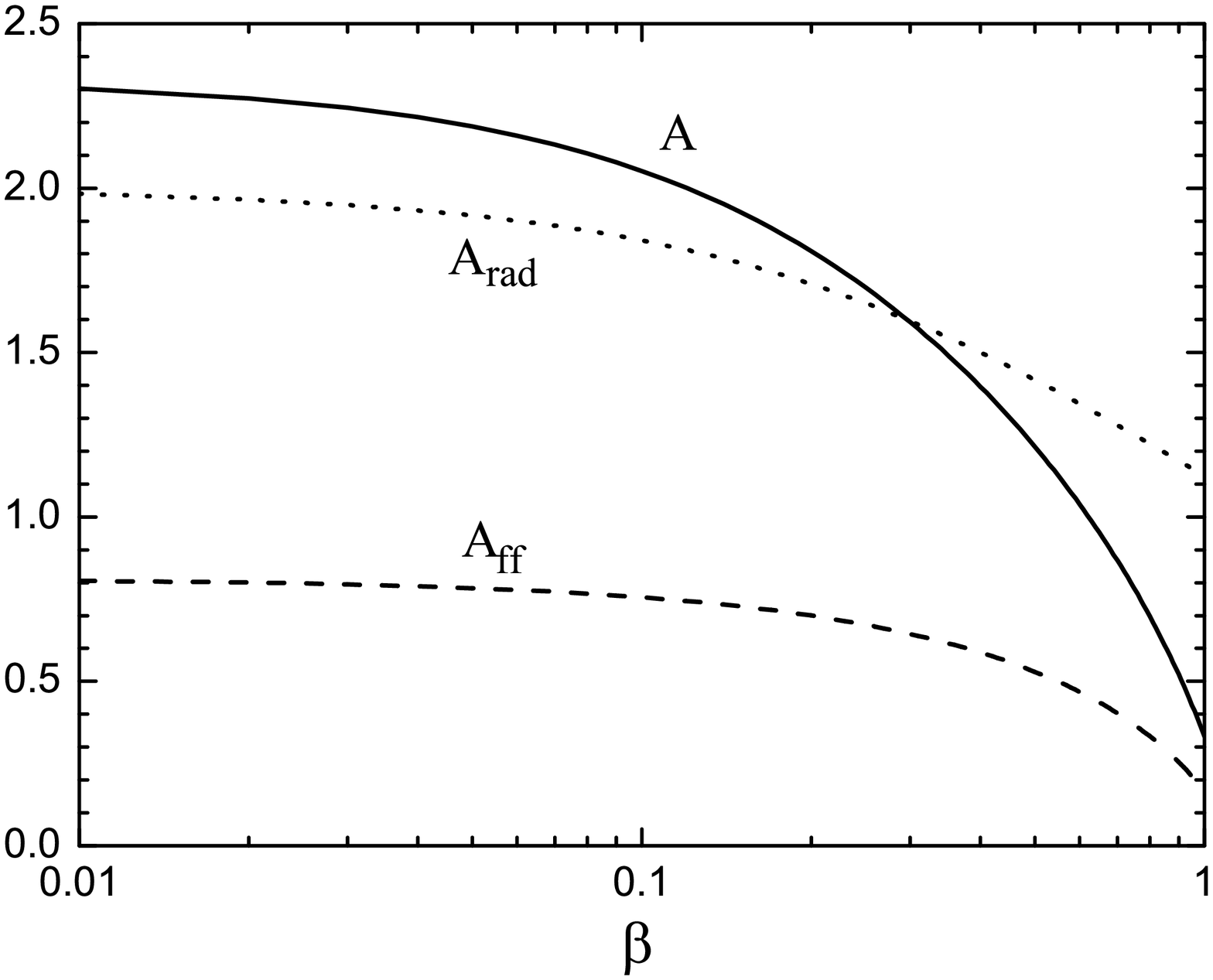}
\caption{
Variations of $A$ (solid line), $A_{\rm rad}$ (dotted line),
and $A_{\rm ff}$ (dashed line) with $\beta$.
\label{Graph1}}
\end{figure}

We use the subscripts ``0" and ``1" to describe
the unperturbed and perturbed quantities, respectively.
We would stress that, the amplitude of fluctuations
is assumed to be small in our linear analysis. In simulations
\citep[e.g., Figures 3 and 4 of][]{Hirose2009b}, however, the amplitudes
can be significantly large.
Nevertheless, the linear analysis may reveal the relationship of fluctuations
of physical quantities.
Combining Equations (\ref{EEnergyE})-(\ref{EState})
with $\bar{\kappa}=\kappa_{\rm es}$, we have
\begin{equation}\label{ETemperatureP}
A{t_{{\rm{th}}}}{\partial  \over {\partial t}}
(4{{{T_1}} \over {{T_0}}}) 
= {{Q_{{\rm{vis,1}}}^ + } \over {Q_{{\rm{vis,0}}}^ + }} - 4{{{T_1}} \over {{T_0}}},
\end{equation}
where the dimensionless parameter $A$ is expressed as
\begin{equation}\label{Afactor}
A=
{{(4 - 3\beta )(\gamma  - 1)} \over {4(1 + \beta )\left[ {\beta  + 3\left( {1 - \beta } \right)(\gamma  - 1)} \right]}}\left[ {7 - 6\beta + {{2\beta } \over {\gamma  - 1}} - {{7\beta \left( {4 - 3\gamma } \right)\left( {1 - \beta } \right)} \over {\left( {\gamma  - 1} \right)\left( {4 - 3\beta } \right)}}} \right],
\end{equation}
and the thermal time-scale $t_{\rm th}$ takes the from (with $Q_{\rm vis,0}^+=3\alpha\Pi_0 \Omega_{\rm K}/2$): 
\begin{equation}\label{ThermalTime-scale}
{t_{{\rm{th}}}} \equiv {{{E_0}} \over {Q_{{\rm{rad}},{\rm{0}}}^ - }} = {{{E_0}} \over {Q_{{\rm{vis}},{\rm{0}}}^ + }} = \left[ {2\left( {1 - \beta } \right) + {2\beta  \over {3(\gamma  - 1)}}} \right]{1 \over {\alpha {\Omega _{\rm{K}}}}}.
\end{equation}
We choose $\gamma =5/3$ and $\alpha =0.02$ (e.g., \citealp{Hirose2009a}) for numerical calculations.
The variation of $A$ with $\beta$ is shown by the solid line
in Figure~\ref{Graph1}.
The other two parameters, $A_{\rm ff}$ (dashed line) and $A_{\rm rad}$
(dotted line), will be introduced by Equations~(\ref{Aff}) and (\ref{Arad}),
respectively.

By assuming that the time-dependent component of fluctuations takes the form
of $\exp (i\omega t)$, e.g.,
${{Q_{{\rm{vis}},{\rm{1}}}^ + }/{Q_{{\rm{vis}},{\rm{0}}}^ + }}$
$\propto$ $\exp(i\omega t)$, we have the following relationship from
Equation~(\ref{ETemperatureP}):
\begin{equation}\label{Erelationship}
{\left( {4{{{T_1}} \over {{T_0}}}} \right)_\omega} = {1 \over {i A\omega{t_{{\rm{th}}}} + 1}}{\left( {{{Q_{{\rm{vis}},{\rm{1}}}^ + } \over {Q_{{\rm{vis}},{\rm{0}}}^ + }}} \right)_\omega},
\end{equation} 
where ${\left( {{{Q_{{\rm{vis}},{\rm{1}}}^ + } /{Q_{{\rm{vis}},{\rm{0}}}^ + }}} \right)_\omega}$ and 
${\left( {4{{{T_1}} / {{T_0}}}} \right)_\omega}$ represent the 
fluctuations with $\omega$ of the viscous heating and
those of the radiative cooling, respectively.
We would stress that Equation (\ref{Erelationship}) is a key relationship
in the present work. 

\section{Comparison of analysis and simulation}

\subsection{Time delay between magnetic energy and radiation energy}

Equation (\ref{Erelationship}) can be modified as
\begin{equation}\label{EE}
{\left( {4{{{T_1}} \over {{T_0}}}} \right)_\omega }
= {1 \over {\sqrt {{{\left( {A\omega {t_{{\rm{th}}}}} \right)}^2} + 1} }}{\left( {{{Q_{{\rm{vis}},{\rm{1}}}^ + } \over {Q_{{\rm{vis}},{\rm{0}}}^ + }}} \right)_\omega }\exp \left( { - i\omega{t_{\rm del}}} \right),
\end{equation}
where the delay time $t_{\rm del}$ of the radiative cooling
compared with the viscous heating takes the form:
\begin{equation}
t_{\rm del} = {1 \over \omega }\arctan \left( {A\omega{t_{{\rm{th}}}}} \right).
\end{equation}
Obviously, this equation implies $t_{\rm del}\approx At_{\rm th}$
for long time-scale fluctuations.

\begin{figure}
\plotone{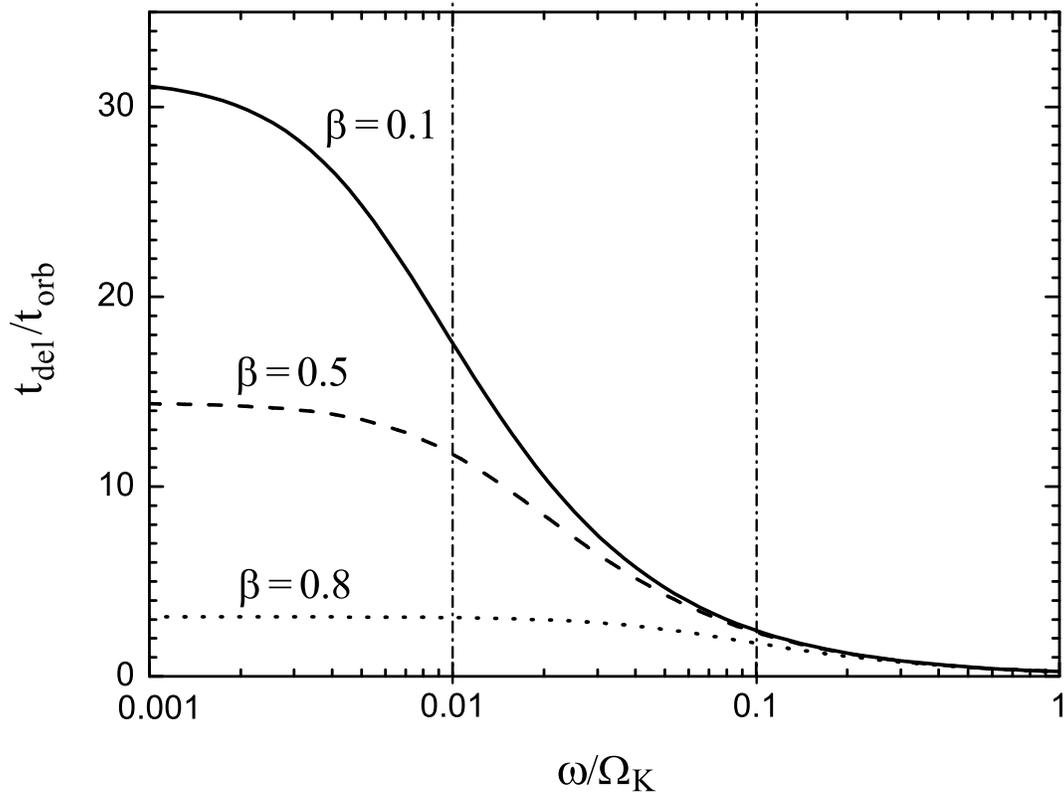}
\caption{
The analytic delay of radiation energy compared with magnetic energy for
$\beta = 0.1$ (solid line), $0.5$ (dashed line), and $0.8$ (dotted line).
The vertical dot-dashed lines represent two specific frequencies
$\omega/\Omega_{\rm K} = 0.01$ and $0.1$.
\label{Graph2}}
\end{figure}

Figure~\ref{Graph2} shows the variation of $t_{\rm del}$
for three different values of $\beta$.
The simulations for $\beta \sim 0.1$ (\citealp{Hirose2009b}) showed that
fluctuations of magnetic energy lead those of radiation energy by
$5-15$ orbit periods ($t_{\rm orb}$), roughly a thermal time.
Moreover, Figure~5 of \citet{Hirose2009b} indicates that
significant variability occurs in the range
$0.01 \la f\cdot t_{\rm orb} \la 0.1$, which is equivalent to
$0.01 \la \omega/\Omega_{\rm K} \la 0.1$, corresponding to the region
between the two vertical dot-dashed lines in Figure~2.
As shown by the solid line, the delay in our analysis is around
$3-17 t_{\rm orb}$, which is consistent with the simulations.
In addition, we would point out that the delay between viscous heating
and magnetic energy ($\sim 0.5t_{\rm orb}$, \citealp{Hirose2009b})
is negligible compared with the thermal time-scale.

Furthermore, simulations have been done for the gas-pressure-dominated
case (\citealp{Hirose2006}) and the case that gas and radiation pressures
are comparable (\citealp{Krolik2007}). The delay in those simulations
is $\sim 2t_{\rm orb}$ for $\beta \sim 0.8$ and $\sim 5t_{\rm orb}$
for $\beta \sim 0.5$. In our analysis, as shown by the dotted and dashed lines, 
the delay is around $2 - 3 t_{\rm orb}$ for $\beta=0.8$ and
$2 - 12 t_{\rm orb}$ for $\beta=0.5$
in the range $0.01 \la \omega/\Omega_{\rm K} \la 0.1$,
which is again consistent with the simulations.
Note that $\bar{\kappa}$ in the simulations for $\beta \sim 0.8$
(\citealp{Hirose2006}) is dominated by the free-free absorption.
In such case, for a simple approach, we modify the parameter $A$ as
$A_{\rm ff}$ by considering the free-free absorption instead of
the electron scattering in Equation~(3):
\begin{equation}\label{Aff}
A_{\rm{ff}} = {{(4 - 3\beta )(\gamma  - 1)} \over {\left( {11.5 + 4.5\beta } \right)\left[ {\beta  + 3\left( {1 - \beta } \right)(\gamma  - 1)} \right]}} \\
\left[ {7 - 6\beta + {{2\beta } \over {\gamma  - 1}} - {{7\beta \left( {4 - 3\gamma } \right)\left( {1 - \beta } \right)} \over {\left( {\gamma  - 1} \right)\left( {4 - 3\beta } \right)}} } \right].
\end{equation}
The profile of $A_{\rm ff}$ is shown by the dashed line in Figure~1.

\subsection{Power spectrum relationship}
\label{PowerSpectralRelationship}

In this subsection, we will show a comparison between our analytic
normalized power spectrum of radiation energy $P_{\rm rad}^{\rm A}$
and that in simulations $P_{\rm rad}$.
Equation~(\ref{EE}) provides the relationship of the power spectrum
between the radiative cooling $P_{\rm cool}$ and the viscous heating
$P_{\rm vis}$:
\begin{equation}\label{Spectral}
P_{\rm cool}(f) 
= {1\over {{1+ (2\pi Af{t_{{\rm{th}}}})^2}}}{P_{\rm vis}}(f).
\end{equation}
Then the analytic power spectrum of volume-integrated radiation energy
$P_{\rm rad}^{\rm A}$ is expressed as
\begin{equation}\label{EPArad}
P_{\rm rad}^{\rm A}(f) = \frac{A_{{\rm{rad}}}^2}{{1 + {{(2\pi Af{t_{{\rm{th}}}})}^2}}}{P_{\rm{vis}}}(f),
\end{equation}
where the quantity $A_{\rm rad}$ is derived from Equations (\ref{EEnergyE}),
(\ref{EVertical}), and (\ref{EState}):
\begin{equation}\label{Arad}
{A_{\rm rad}} \equiv {(E_{\rm rad, 1}/E_{\rm rad,0})\over (4{{{T_1}}/{{T_0}}}) }= 1 + \frac{{4 - 3\beta}}{{4(1 + \beta )}}.
\end{equation}
The variation of $A_{\rm rad}$ with $\beta$
is shown by the dotted line in Figure~\ref{Graph1}. 
Since the viscous heating is mainly due to the dissipation of
magnetic energy, it is plausible to have
\begin{equation}\label{PmPvis}
{P_{{\rm{vis}}}}(f) \approx \left [{(\sqrt{2}\pi f{t_{{\rm{dis}}}})^2} + 1 \right]{P_B}(f),
\end{equation}
where $t_{\rm dis}$ is the dissipation time-scale of magnetic energy,
and $P_B$ is the power spectrum of magnetic energy.
The explanation for this relationship is presented in Appendix A.
Then, Equations (\ref{EPArad}) and (\ref{PmPvis}) provide
an analytic relationship between the power spectrum of radiation energy
and that of magnetic energy:
\begin{equation}\label{EPvisPB}
P_{\rm rad}^{\rm A}(f) = A_{{\rm{rad}}}^2\frac{{1 + {{(\sqrt{2}\pi f{t_{{\rm{dis}}}})}^2}}}{{1 + {{(2\pi Af{t_{{\rm{th}}}})}^2}}}{P_B}(f).
\end{equation}

The simulations \citep[Equation~(17) and Figure~5 of][]{Hirose2009b}
showed the profiles of $P_B$ and $P_{\rm rad}$:
\begin{equation}
{P_B}\left( f \right) = \left\{ {\matrix{
   {8.7 \times {{10}^{ - 6}}{f^{ - 1.13}},} & {f < 0.171,}  \cr 
   {{{10}^{ - 7}}{f^{ - 3.65}},} & {f > 0.171,}  \cr 
 } } \right.
\end{equation}
\begin{equation}
P_{\rm rad} \left( f \right) = \left\{ {\matrix{
   {6.1 \times {{10}^{ - 9}}{f^{ - 2.38}},} & {f < 0.118,}  \cr 
   {2.3 \times {{10}^{ - 10}}{f^{ - 3.91}},} & {f > 0.118.}  \cr 
 } } \right.
\end{equation}
In our Figure~3, we replot the above $P_B$ and $P_{\rm rad}$
with the dotted and dashed lines, respectively. In addition,
according to Equation~(\ref{EPvisPB}), we plot the analytic power spectrum
$P_{\rm rad}^{\rm A}$ with the solid line. The values of $P_B (f)$
in Equation~(\ref{EPvisPB}) are taken from the above simulation results
(the dotted line).
The parameters for calculating $P_{\rm rad}^{\rm A}$ are
$A=2.1$, $A_{\rm rad}=1.8$ (corresponding to $\beta=0.1$), and
$t_{\rm dis}=0.5t_{\rm orb}$ (\citealp{Hirose2009b}).
As shown by the solid and dashed lines, our analytic
$P_{\rm rad}^{\rm A}$ agrees well with $P_{\rm rad}$
in simulations.

\subsection{Correlation between magnetic energy and radiation energy}

The energy equation implies a correlation between the viscous heating
(magnetic energy) and the pressure. 
Based on Equation~(\ref{ETemperatureP}), the correlation can be read as:
\begin{equation}
{{Q_{{\rm{vis,1}}}^ + } \over {Q_{{\rm{vis,0}}}^ + }} \simeq 4{{{T_1}} \over {{T_0}}}.
\end{equation}
Owning to the same reason, there should be a correlation 
between viscous heating and magnetic energy, i.e., 
$Q_{\rm{vis,1}}^+ / Q_{\rm{vis,0}}^+ \simeq E_{B,1} / E_{B,0}$.
With Equation~(16), the above equation can be modified as
\begin{equation}\label{E22}
{{E_{B,1}} \over {E_{B,0}}}
\simeq {1\over A_{\rm rad}}{{{E_{\rm rad,1}}} \over {{E_{\rm rad,0}}}}.
\end{equation}
As shown in Figure~1, there exists $A_{\rm rad} = 1.8$ for $\beta=0.1$,
so we have the relationship:
\begin{equation}\label{ECorrelation}
E_B \propto E_{\rm rad}^{0.55},
\end{equation}
which is close to the correlation found in simulations, e.g., 
$E_B \propto \; E_{\rm rad}^{0.71}$ (\citealp{Hirose2009b}).
The difference in the index may be related to the following two
reasons: (1) the analysis is quite simple, particularly in dealing
with the vertical radiative cooling; (2) the feedback from pressure
to stress makes significant contribution.

It is worthy to note that \cite{Hirose2009b} also provided an explanation
for the correlation with a toy model based on the energy equation.
By using the correlation between $t_{\rm th}$ and $E_{\rm rad}$
($t_{\rm th} \propto E_{\rm rad}^s$) obtained in simulations, 
they derived the correlation between $E_B$
and $E_{\rm rad}$ ($E_B \propto E_{\rm rad}^{1-s}$).
In our analysis, we choose $Q_{\rm rad}^-$ to replace
their radiative cooling term $E_{\rm rad}/t_{\rm th}$.
We will show below that our results are quite similar to theirs.

Equations (\ref{ERadiationCooling}) and (\ref{Arad}) can provide the
relationship:
\begin{equation}\label{ECorrelationQrad}
Q_{\rm rad}^- \propto E_{\rm rad}^{1/A_{\rm rad}}.
\end{equation}
In simulations (\citealp{Hirose2009b}), the thermal time is calculated by
\begin{equation}
t_{\rm th}=E/Q_{\rm rad}^-.
\end{equation}
With Equations (\ref{EEnergyE}),
(\ref{EVertical}), (\ref{EState}) and (\ref{ECorrelationQrad}), 
the above equation can be reduced to 
\begin{equation}
t_{\rm th}\propto E_{\rm rad}^{s'},
\end{equation}
where 
\begin{equation}\label{Es}
s' =
\frac{{3(1 - \beta )(4 - 3\beta )(\gamma  - 1) - 3\beta (1 + \beta )}}
{{(8 + \beta )[\beta  + 3(\gamma  - 1)(1 - \beta )]}}.
\end{equation}
For radiation-pressure-dominated accretion flows,
Equation (\ref{Es}) can be simplified as $s' \approx 1-1 / A_{\rm rad}$.
Thus, Equation (\ref{E22}) indicates the relationship
$E_B \propto E_{\rm rad}^{1-s'}$, which is
consistent with the correlation in the toy model \citep{Hirose2009b}.
Moreover, in the case of $\beta=0.1$, Equation (\ref{Es}) gives $s'=0.41$,
thus $t_{\rm th} \propto E_{\rm rad}^{0.41}$,
which is close to the correlation found in simulations, 
e.g., $t_{\rm th} \propto E_{\rm rad}^{0.32}$ in simulation 1112a and 
$t_{\rm th} \propto E_{\rm rad}^{0.44}$ in simulation 1126b
of \cite{Hirose2009b}.

\section{Conclusions and Discussion}
In the present work, we have studied the fluctuations of standard
thin disks by linear analysis of the time-dependent
energy equation together with the vertical hydrostatic 
equilibrium and the equation of state.
Our analytic results show that the delay between magnetic energy
and radiation energy is consistent with that in previous simulations.
In addition, the analytic power spectrum of radiation energy
agrees well with that in simulations.
Moreover, the correlation between magnetic energy and radiation energy
can be well understood by the analysis, with an index ($0.55$) being
close to that in simulations ($0.71$).

As indicated by Equation~(\ref{Spectral}), there may exist a break frequency
$f_{\rm br}\sim 1/(2{\pi}A{t_{{\rm{th}}}})$ in $P_{\rm cool}$.
The frequency $f_{\rm br}$ may be associated with the high-frequency break 
observed in the power spectra of luminosity fluctuations
(e.g., \citealp{McHardy10}), 
since its value in the inner region of disk is close to that of observed high frequency break.
Moreover, the difference between $P_{\rm cool}$ and $P_{\rm vis}$
for $f>f_{\rm br}$, shown by Equation (\ref{Spectral}), should be taken into account in 
modeling the high-frequency variability of quasar luminosity
(e.g., \citealp{Mushotzky2011}; \citealp{Zu2012}).
In addition, the similar frequency break may also occur in
radiatively inefficient accretion flows,
such as advection-dominated accretion flows (\citealp{Narayan1994})
and slim disks (\citealp{Abramowicz1988}).

\acknowledgments
We thank the referee, Omer Blaes, for helpful suggestions and
useful communications to improve the paper.
We also thank Feng Yuan and Sheng-Ming Zheng for beneficial discussions.
This work was supported by the National Basic Research Program (973 Program)
of China under grant 2009CB824800, and the National Natural Science Foundation
of China under grants 10833002, 11073015, 11103015, 11222328, and 11233006.

\appendix
\section{The relationship between $P_{\rm vis}$ and $P_B$}
In this Appendix, we try to derive the relationship
between $P_{\rm vis}(f)$ and $P_B(f)$ as shown by Equation (\ref{PmPvis}).
The evolution of magnetic energy $E_B(t)$ can be simply described as 
\begin{equation}\label{EMagneticEnergyEvolution}
{\partial {E_B}(t) \over \partial t}= G_B(t)  - D_B(t),
\end{equation}
where $G_B(t)$ and $D_B(t)$ are respectively the generation 
and dissipation rate of magnetic energy. 
With small amplitude perturbations in Equation (\ref{EMagneticEnergyEvolution}), 
we have  
\begin{equation}
i\omega t_{\rm dis}\left({E_{B,1} \over E_{B,0}}\right)_{\omega} 
= \left({G_{B,1} \over G_{B,0}}\right)_{\omega} - 
\left({D_{B,1} \over D_{B,0}}\right)_{\omega},
\end{equation}
or 
\begin{equation}\label{EPowerSpectraRelationship}
{\left( {\omega t_{\rm dis}} \right)}^2 {P_B(\omega)} = {P_G(\omega)} + {P_D(\omega)} - {\left( {{{{G_{B,1}}} \over {{G_{B,0}}}}} \right)_\omega } \times {\left[ {{{\left( {{{{D_{B,1}}} \over {{D_{B,0}}}}} \right)}_\omega }} \right]^{\ast}} - {\left[ {{{\left( {{{{G_{B,1}}} \over {{G_{B,0}}}}} \right)}_\omega }} \right]^{\ast}} \times {\left( {{{{D_{B,1}}} \over {{D_{B,0}}}}} \right)_\omega },
\end{equation}
where $D_{B,0}=G_{B,0}$, $t_{\rm dis}=E_{B,0}/D_{B,0}$,
the symbol ``$\ast$'' represents the complex conjugate number,
the power spectrum of $G_B$ and $D_B$ are

\[
P_G(\omega)={\left| {{{\left( {{{{G_{B,1}}} \over {{G_{B,0}}}}} \right)}_\omega }} \right|^2},\;\;
P_D(\omega)={\left| {{{\left( {{{{D_{B,1}}} \over {{D_{B,0}}}}} \right)}_\omega }} \right|^2}.
\]

Magnetic fields in the accretion disk present exponential growth
owing to the magnetorotational instability (MRI, \citealp{Balbus1991}),
followed by the dissipation due to some destructive mechanisms. 
Since the rise and decay phases of channel modes are
similar (e.g., Figure~4 of \citealp{Simon2009}), 
it is plausible to believe that $P_G$ and $P_D$ is 
comparable for $\omega \lesssim \pi/t_{\rm dis}$.
With the consideration of the delay ($\sim t_{\rm dis}$) between $G_B$ and $D_B$, 
the relationship between $G_B$ and $D_B$ can be modeled as 
\begin{equation}
{\left( {{{{D_{B,1}}} \over {{D_{B,0}}}}} \right)_\omega } 
\sim {\left( {{{{G_{B,1}}} \over {{G_{B,0}}}}} \right)_\omega }\exp \left( { - i\omega {t_{\rm dis}}} \right).
\end{equation}
Substituting this relationship into Equation
(\ref{EPowerSpectraRelationship}), we obtain
\begin{equation}\label{EPSDLongTime}
{P_D(\omega)} = {{{\left( {\omega t_{\rm dis}} \right)}^2} \over 
{2\left( {1 - \cos \omega {t_{\rm dis}}} \right)}}{P_B(\omega)}
\approx \left [{(\omega t_{\rm dis})^2 \over 2} + 1 \right ]{P_B}(f)
.
\end{equation}
The above relationship is applicable for $\omega \lesssim \pi/t_{\rm dis}$.

The fluctuations with $\omega > \pi/t_{\rm dis}$ in $G_B$ and those in $D_B$
is unclear, 
and thus it remains uncertain for $P_G(\omega)$ and $P_D(\omega)$. 
However, it may be plausible to believe that $G_B$ and $D_B$ 
are decoupled with each other for $\omega > \pi/t_{\rm dis}$.
If we further assume that $P_G(\omega) \sim P_D(\omega)$,
Equation (\ref{EPowerSpectraRelationship}) can be reduced to 
\begin{equation}\label{EPSDShortTime}
{P_D}(\omega) \sim {(\omega {t_{{\rm{dis}}}})^2 \over 2}{P_B}(\omega).
\end{equation}
We use this equation to describe the relationship of 
${P_D}(\omega)$ and ${P_B}(\omega)$ for $\omega > \pi/t_{\rm dis}$. 
Based on Equations (\ref{EPSDLongTime}) and (\ref{EPSDShortTime}), 
a general form of relationship between $P_D$ and $P_B$ 
may be simply described as  
\begin{equation}
{P_D}(f) \approx \left [{(\sqrt{2}\pi f{t_{{\rm{dis}}}})^2} + 1 \right ]{P_B}(f).
\end{equation}
Since the turbulent kinetic energy follows the 
fluctuating magnetic energy (\citealp{Hirose2009b})
and magnetic dissipation dominates over kinetic dissipation,
we obtain 
\begin{equation}
{P_{{\rm{vis}}}}(f) \approx {P_D}(f) \approx \left [{(\sqrt{2}\pi f{t_{{\rm{dis}}}})^2} + 1 \right ]{P_B}(f),
\end{equation}
which is the exact form of Equation~(\ref{PmPvis}). 
It should be noted that the relationship between $P_{\rm vis}$ and $P_B$ 
in the short time-scale range ($\omega > \pi/t_{\rm dis}$) is tentatively used in the present work.

\clearpage

\clearpage

\clearpage

\begin{figure}
\plotone{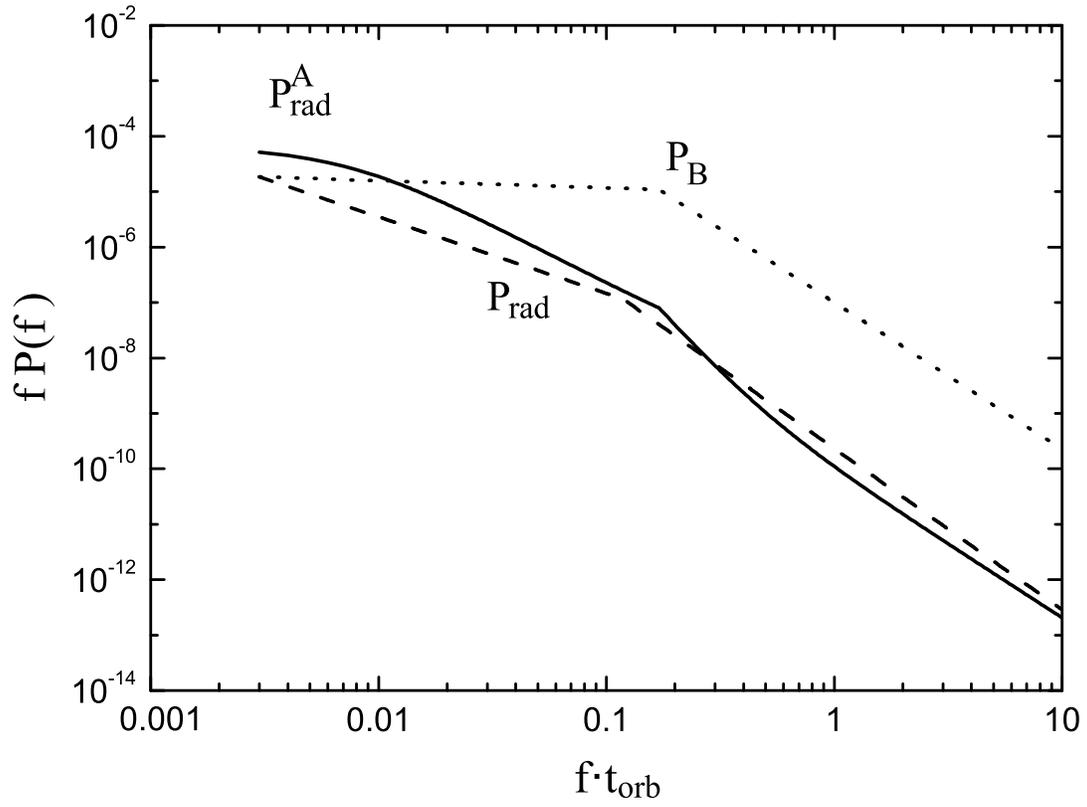}
\caption{
A comparison of $P_{\rm rad}^{\rm A}$ (solid line) and $P_{\rm rad}$
(dashed line), where $P_{\rm rad}^{\rm A}$ is calculated
with $P_B$ (dotted line).
\label{Graph3}}
\end{figure}


\begin{thebibliography}{} 
\bibitem[Abramowicz et al.(1988)]{Abramowicz1988}
    Abramowicz, M.~A., Czerny, B., Lasota, J.~P., \& Szuszkiewicz, E.\ 1988, \apj, 332, 646 
\bibitem[Balbus \& Hawley(1991)]{Balbus1991}
    Balbus, S.~A., \& Hawley, J.~F.\ 1991, \apj, 376, 214
\bibitem[Ciesielski et al.(2012)]{Ciesielski2012}
    Ciesielski, A., Wielgus, M., Klu{\'z}niak, W., et al.\ 2012, \aap, 538, A148 
\bibitem[Hawley \& Krolik(2001)]{HawleyK2001}
    Hawley, J.~F., \& Krolik, J.~H.\ 2001, \apj, 548, 348 
\bibitem[Hirose et al.(2009a)]{Hirose2009a}
    Hirose, S., Blaes, O., \& Krolik, J.~H.\ 2009a, \apj, 704, 781
\bibitem[Hirose et al.(2009b)]{Hirose2009b}
    Hirose, S., Krolik, J.~H., \& Blaes, O.\ 2009b, \apj, 691, 16 
\bibitem[Hirose et al.(2006)]{Hirose2006}
    Hirose, S., Krolik, J.~H., \& Stone, J.~M.\ 2006, \apj, 640, 901 
\bibitem[Kato et al.(2008)]{Kato2008}
    Kato, S., Fukue J., \& Mineshige, S.\ 2008, Black-Hole Accretion Disks: 
    Toward a New Paradigm (Kyoto: Kyoto Univ. Press)
\bibitem[King et al.(2004)]{King2004}
    King, A.~R., Pringle, J.~E., West, R.~G., \& Livio, M.\ 2004, \mnras, 348, 111 
\bibitem[Krolik et al.(2007)]{Krolik2007}
    Krolik, J.~H., Hirose, S., \& Blaes, O.\ 2007, \apj, 664, 1045
\bibitem[Lightman \& Eardley(1974)]{Lightman1974}
    Lightman, A.~P., \& Eardley, D.~M.\ 1974, \apj, 187, L1
\bibitem[Lin et al.(2011)]{Lin2011}
    Lin, D.-B., Gu, W.-M., \& Lu, J.-F. 2011, \mnras, 415, 2319
\bibitem[Lyubarskii(1997)]{Lyubarskii1997}
    Lyubarskii, Y.~E.\ 1997, \mnras, 292, 679 
\bibitem[McHardy(2010)]{McHardy10}
    McHardy, I.\ 2010, in The Jet Paradigm, ed. T. Belloni 
    (Lecture Notes in Physics, Vol. 794; Berlin: Springer), 203
\bibitem[Mushotzky et al.(2011)]{Mushotzky2011}
    Mushotzky, R.~F., Edelson, R., Baumgartner, W., \& Gandhi, P.\ 2011, \apj, 743, L12 
\bibitem[Narayan \& Yi(1994)]{Narayan1994}
    Narayan, R., \& Yi, I. 1994, \apj, 428, L13
\bibitem[Noble \& Krolik(2009)]{Noble2009}
    Noble, S.~C., \& Krolik, J.~H.\ 2009, \apj, 703, 964 
\bibitem[Shakura \& Sunyaev(1973)]{Shakura1973}
    Shakura, N.~I., \& Sunyaev, R.~A.\ 1973, \aap, 24, 337 
\bibitem[Shakura \& Sunyaev(1973)]{Shakura1976}
    Shakura, N.~I., \& Sunyaev, R.~A.\ 1976, \mnras, 175, 613 
\bibitem[Simon et al.(2009)]{Simon2009}
    Simon, J.~B., Hawley, J.~F., \& Beckwith, K.\ 2009, \apj, 690, 974 
\bibitem[Zu et al.(2012)]{Zu2012} 
    Zu, Y., Kochanek, C.~S., Koz{\l}owski, S., \& Udalski, A.\ 2012, arXiv:1202.3783
\end{thebibliography}
\end{document}